
\def\refset{\parindent=0pt\hangafter=1\hangindent=1em}
\def\etal{{et al.}\ }
\def\cf{{cf.}\ }

\def\kms{\rm km s^{-1}}
\def\mpc{\rm Mpc }

\magnification=1200
\def\lap{\lower.5ex\hbox{$\; \buildrel < \over \sim \;$}}
\def\gap{\lower.5ex\hbox{$\; \buildrel > \over \sim \;$}}
\parskip 3pt plus 1pt minus .5pt
\baselineskip 19pt plus .1pt
\newcount\eqtno
\eqtno = 1
\font\eightrm=cmr8 scaled \magstep0
\def\fig #1, #2, #3, #4, #5, #6 {
\topinsert
\smallskip
\centerline{\psfig{figure=#1,height=#2 in,width=#3 in,angle=#4}}
\medskip
{\vskip #5 cm\leftskip2.5em \parindent=0pt {\eightrm #6 }}
\endinsert}
%
%
\centerline{\ \ }
\vskip 0.5in
\centerline{\bf TOPOLOGY OF LARGE-SCALE STRUCTURE BY GALAXY TYPE:}
\centerline{\bf HYDRODYNAMIC SIMULATIONS}
\bigskip
\bigskip
\centerline{J. Richard Gott, III, Renyue Cen, and Jeremiah P. Ostriker}
\vskip 1.0cm
\centerline{Princeton University Observatory}
\centerline{Princeton, NJ 08544}
\vskip 0.5in
\centerline{email: cen@astro.princeton.edu}
\vskip 0.5in
\centerline{Submitted to {\it The Astrophysical Journal}}
\vskip 0.3in
\centerline{March 24, 1995}
\vfill\eject

\centerline{ABSTRACT}

The topology of large scale structure is studied as
a function of galaxy type using
the genus statistic.  In hydrodynamical cosmological
CDM simulations, galaxies form on
caustic surfaces (Zeldovich pancakes) then slowly
drain onto filaments and clusters.  The
earliest forming galaxies in the simulations
(defined as ``ellipticals") are thus seen at the present epoch
preferentially in clusters (tending toward a meatball topology),
 while the latest forming
galaxies (defined as ``spirals")
 are seen currently in a spongelike topology.  The topology is measured by
the genus (= number of ``donut" holes - number of
isolated regions) of the smoothed
density-contour surfaces.  The measured genus curve
 for all galaxies as a function of
density obeys approximately the theoretical curve
expected for random-phase initial
conditions, but the early forming elliptical galaxies
 show a shift toward a meatball topology
relative to the late forming spirals.
Simulations using standard biasing schemes fail to
show such an effect.  Large observational
samples separated by galaxy type could be used
to test for this effect.

\noindent
Cosmology: large-scale structure of Universe
-- galaxies: clustering
-- galaxies: formation
-- hydrodynamics
-- stars: formation
\vfill\eject

\centerline{1. INTRODUCTION}

	The formation of large-scale structure
in the universe remains one of the most
interesting problems in cosmology.  Much progress
in understanding this question has
been made in the last quarter century.
The standard picture for structure formation has been
the gravitational instability picture:
small random-phase initial fluctuations growing under
the action of gravity (for discussions, see Gunn \& Gott 1972;
Doroshkevich, Zeldovich \& Sunyaev 1976;
Peebles 1980).
This picture predicted an essentially undistorted thermal
spectrum for the cosmic microwave background
(confirmed by the COBE satellite, \cf Mather et al. 1990),
and small angular fluctuations in the cosmic microwave temperature
(Sachs \& Wolfe 1967), also
confirmed by the COBE satellite (Smoot et al. 1992).  Models where the mass
in the universe is dominated by Cold Dark Matter (CDM) (Peebles 1982)
(with $\Omega$ not necessarily unity)
have been   particularly successful in explaining a number of observations
(cf. White \etal 1987; Park 1990; Park \& Gott 1991;
Weinberg \& Gunn 1990),
in particular, the order of magnitude of the amplitude of the COBE
fluctuations and the observed pattern of great walls
and voids seen in the data (de
Lapparent, Geller \& Huchra 1986; Geller \& Huchra 1989).
A review of the virtues and defects of the CDM models
is presented in Ostriker (1993).
While the standard ($\Omega_{CDM}=1$) model is almost certainly
not correct, the best fitting models  now under discussion
are variants of standard CDM,
which retain many of the essential features of the original model.
In CDM models, caustics form
(Zeldovich Pancakes, cf. Doroshkevich, Zeldovich, \& Sunyaev 1976;
Einasto, Joeveer \& Saar 1980)
and
thus Great Walls are produced naturally
(cf. Park 1990;
Weinberg \& Gunn 1990;
Park \& Gott 1991).
As such models evolve in time,
there is a draining of matter by
gravity from minor walls onto filaments which produces
 an even larger network of voids as
minor walls become so thinly populated as to
effectively disappear (Weinberg 1990).  This
leaves a distinctive pattern of major
and minor walls.  The observational sample shows
such features as well.
In the original de Lapparent et al. $6^o$-wide slice, the large elliptical
void on the left is nearly bisected by a very sparse and very thin wall of
galaxies (Park et al.
1992a) which was noticed
in a topology study and which looks just like the minor walls
seen in the simulations.
Deep pencil beams to redshift $z = 0.5$ show great walls with a
median separation of
$133 h^{-1}$\mpc in the CDM simulations (Park \& Gott 1991) and
$128 h^{-1}$\mpc in the observations (Broadhurst et al. 1990).
Broadhurst et al. reported a regular
periodicity for these walls as measured by a power spectrum test,
but 1 of the 12 simulated
pencil beams showed an even greater periodicity by the same test (Park \& Gott
1991), so
the Broadhurst et al. results were certainly consistent with the standard model
at the $2\sigma$ level.
The inflationary CDM spectrum (Bardeen et al. 1986),
gives an excellent fit to the
large scale power seen in a variety of samples (Maddox et al. 1990; Saunders et
al. 1991
[IRAS]; Shectman et al. 1995 [deep slices]; Park, Gott \& da Costa 1992
[Southern Sky
Survey]) providing that
$\Omega h \sim 0.3$ where
$h = (H_o/100 \kms \mpc^{-1})$.
Given current estimates of the Hubble constant of $h \sim 0.8$
(Freedman et al. 1994; Pierce et al. 1994), this
would seem to favor either low-density ($\Omega <1$),
single-bubble inflationary models (cf. Gott 1982;
Gott \& Statler 1984; Gott 1986; Bucher, Goldhaber \& Turok 1995;
Yamamoto, Sasaki, \&
Tanaka 1995; Ratra \& Peebles 1994; Kamionkowski \& Spergel 1994)
or $k = 0$, $\Omega_{CDM}\sim 0.4$, $\Omega_\Lambda \sim 0.6$ inflationary
model
(cf.
Bahcall \& Cen 1992;
Efstathiou, Bond, \& White 1992;
Kofman, Gnedin, \& Bahcall 1993;
Cen, Gnedin, and Ostriker 1993;
Cen \& Ostriker 1994).
Alternatively
$\Omega_{CDM} = 1$, $h = 0.3$ models
(Bartlett, Blanchard, Silk, \& Turner 1995),
 Mixed (70\% cold + 30\% hot)
Dark Matter Models (Davis, Summers \& Schegel 1992; Taylor \& Rowan-
Robinson 1992; Klypin et al. 1993; Cen \& Ostriker 1994;
Ma \& Bertschinger 1994),
or tilted inflationary power spectra (Cen \etal 1993) have also been proposed.
While standard biased CDM
simulations
(which do not match the COBE normalization)
produce excellent fits to the observations in many ways,
they simply designate proto-galaxies using the expected number density
of galaxy-sized peaks in the CDM
distribution found in the initial conditions and
ignore the important gas dynamical and atomic
radiative effects
that must surely be important at small scales.
Thus Cen \& Ostriker (1992a) have embarked
on a series of hydrodynamical simulations to model these effects
allowing for as much of the relevant physics as possible.

	These hydrodynamical simulations have given promising
results in a number of ways.  As the density fluctuations grow, the gas
dynamics are followed including shock
heating from convergent flows,
subsequent cooling etc.  If there is a collapsing flow and the
local cooling time is shorter than the local collapse time,
one expects that the gas can cool to
permit star formation and
a proto-galaxy is formed.
Proto-galaxies are then followed via collisionless dynamics and
they are grouped
into galaxies (via a friend-of-friend algorithm) at each epoch
only to examine their properties (we do not group them during
the evolution).
Galaxies form on caustics (Zeldovich pancakes = walls) and
filaments - then drain via gravity onto clusters.
If the first 25\% of galaxies formed are
identified as ``ellipticals", the second 25\% as ``S0"
galaxies, and the last 50\% as ``spirals",
a convenient and simple if crude representation of the
observed morphology-mean age relation, then
the morphology-density relation observed
by Dressler (1980) and Bhavsar (1981) is
reproduced well (Cen \& Ostriker 1993a, Figure 6a),
since ellipticals have migrated toward clusters
while spirals are still on a filamentary net (see Fig. 1 panels
a and c).
Also, the elliptical galaxies
produced have a higher covariance function among
themselves than spiral galaxies have among
themselves (Cen \& Ostriker 1993a, Figure 7b),
also in agreement with the observations (cf. Davis \&
Geller 1976).  The distribution of hot and cold intergalactic gas produced in
the simulations
is good, giving approximately the correct number of X-ray clusters (Kang et al.
1994) and
approximately the correct spectrum for the X-ray background
(Cen \etal 1995).
The distribution of Lyman alpha clouds is surprisingly realistic (Cen et al.
1994).

	Now the morphology-density relation of Dressler (1980) and Bhavsar (1981) may
be explained by a variety of reasons.  Elliptical galaxies may simply be those
with the
shortest collapse times as suggested by Gott \& Thuan (1977).  They then have
the highest
mass densities at turn-around and the highest ratios of collapse times to
cooling times and
star formation times, so that they are also the galaxies that complete their
star formation
before  their collapse (cf. Ostriker \& Rees 1977).  S0 and spiral galaxies do
not complete
their star formation before their collapse, so in addition to a spheroidal
component of stars
they have some left-over gas which infalls after the spheroid is formed and
dissipates into a
disk.  Further star formation can then occur in the disk, either reaching
completion (as in
the S0's) or not (as in the spirals).  Spirals subsequently infalling into
clusters  filled with
hot virialized intergalactic gas can be stripped by
ram pressure stripping
(Gott \& Gunn 1971;
Gunn \& Gott 1972)
and turned into S0's.
This would explain the complete absence
of spirals from the cluster cores of great (regular) virialized clusters like
Coma (which are
also X-ray clusters).  Since spirals and S0's have small spheroidal components,
they
would look like small ellipticals if their infall gas did not exist.  Thus, an
elliptical galaxy
with a total luminosity of L* might have a larger total mass (including its CDM
heavy halo),
than an L* luminosity spiral.  Now the 3-point function for galaxy clustering
(Groth \&
Peebles 1977) [with $Q\sim 1$ as observed]
 shows that the covariance function of a tight binary
galaxy with all galaxies should be just twice the amplitude of the
galaxy-galaxy covariance
function (Gott 1980).  This is reasonable since an object that is twice as
massive should
have twice the infall toward it, and a tight binary is dynamically equivalent
to a single object
of twice the total mass.
 Indeed, Gott, Turner, \& Aarseth (1979) did a simulation with
point particles of mass
2M* representing elliptical and S0 galaxies (with their lack of young
bright stars) and particles of mass
M* representing spiral galaxies and found
$\xi_{E,S0-gal} \sim 2\xi_{gal-gal}$
just as in the observations (Davis \& Geller 1976).
In the quoted hydrodynamical
simulations, the elliptical galaxies formed, for example, have an average mass
that is about 3 times as large as the average mass in the spirals.
Also, if
elliptical galaxies are those which collapse first, they represent higher
initial amplitude
peaks in the initial conditions, and according to standard biased galaxy
formation schemes,
they would represent peaks with a higher required threshold than other
galaxies, and
therefore would constitute a more biased initial distribution with a higher
amplitude
covariance function in the initial conditions (Bardeen \etal 1986).

	Thus, there are a number of effects that would independently or together tend
to act
to produce a morphology-density relation similar to that observed.  The
hydrodynamical
simulations naturally produce a distribution of galaxies that is biased
relative to the CDM
mass distribution:
$(\delta \rho_{gal}/\bar\rho_{gal}) \sim b(\delta\rho/\bar\rho)$ where
$b\sim 1.3$ on a scale of $8h^{-1}$\mpc.
 This occurs by natural processes
occurring as the gas turns into galaxies and explains why a standard biasing
scheme of
picking initial peaks in the initial conditions might work fairly
well in practice.
The hydrodynamic simulations that use real physics thus offer us a
good reason why
we might expect
the galaxy distribution to be somewhat biased relative to the mass
distribution.  In this paper we are looking for some additional testable
telltale signs that the
observed distribution of galaxies is in fact being produced by the detailed
mechanisms
actually occurring in the hydrodynamic simulations rather than just those
generic to any
model possessing a biased galaxy distribution.
Topology seems to possess this property,
for if the spirals are truly formed and
found on a filamentary net (which would have a spongelike
topology) and the ellipticals are preferentially found in clusters
(a meatball topology) although
perhaps also occasionally found on filaments
(\cf Cen \& Ostriker 1993a), then using the genus curve
might permit differentiation between the two distributions in a
large observational sample.
This could provide a test of the detailed processes that are
occurring in the hydrodynamical simulations.  In this paper we show how such a
topological analysis can be done and how its results differ when applied to the
hydrodynamical simulations and simple but naive biasing models.  Also we
indicate how
these techniques can be applied to large observational samples in the future.

\medskip
\centerline{2.  Hydrodynamic Simulations}
\medskip
	The numerical methods and treatment of detailed atomic physics are described
in
Cen (1992) and Cen \& Ostriker (1992).  To summarize briefly, we use a
hydrocode based
on Jameson's (1989) aerospace code, modified extensively for cosmological
applications.
Poisson's equation is solved on a periodic mesh with a FFT routine.  All the
principal line
and continuum atomic processes are computed for each cell and each time step
assuming a
plasma of standard primordial composition.  The radiation field from 1 eV to
100 keV is
treated in detail with allowance for sources, sinks and cosmological effects,
but only in a
spatially averaged fashion.

	We model galaxy formation in a heuristic but hopefully plausible way.  The
details
of how we identify galaxy formation, follow the motions of formed galaxies and
treat
feedback processes (UV and supernovae energy input into the IGM from young
massive
stars) have been presented (Cen \& Ostriker 1993a).
First we
check cells for baryonic overdensity and only examine further those with
$(\delta\rho/\bar\rho) > 4.5$
as candidates for regions within which galaxy formation will occur.
Then, of these cells, we
tag those that satisfy the following criteria:
$$\eqalignno{\nabla\cdot{\vec v}<&~0 {\ \ }\qquad\qquad\qquad\Longrightarrow
\hbox{\ \ contraction}\qquad, &(\the\eqtno) \cr}$$
\advance\eqtno by 1
$$\eqalignno{t_{cool} <&~t_{dyn} \equiv \sqrt{3\pi\over 32G\rho_{tot}}
{\ \ }\qquad\qquad\qquad\Longrightarrow\hbox{\ \ cooling rapidly}\qquad,
&(\the\eqtno) \cr}$$
\advance\eqtno by 1
$$\eqalignno{m_B >& m_J \equiv
G^{-3/2}\rho_b^{-1/2}C^3[1+({\delta\rho_d/{\bar\rho}_d\over\delta\rho_b/{\bar\rho}_b})({{\bar\rho}_d\over{\bar\rho}_b})]^{-3/2}{\ \ }\qquad\qquad\qquad\cr
&\Longrightarrow\hbox{\ \ gravitationally unstable}\qquad. &(\the\eqtno) \cr}$$
\advance\eqtno by 1
Here ($\rho_b,\rho_d$) are baryonic and dark matter densities,
($m_b,m_J$) baryonic mass in the cell
and Jeans mass of the cell,
$C$ isothermal sound speed and other
symbols have their usual meanings.
If all of these criteria are satisfied,
it seems impossible
to prevent collapse of the gas towards the center of the cell with subsequent
condensation into a stellar system. This, of course,
we cannot follow with our code.
Instead we adopt a model inspired
by the classic work of Eggen, Lynden-Bell \& Sandage (1962, ELS)
and assume that the dynamical free-fall and galaxy formation timescales
are simply related.
We remove from the gas in the cell
in question the mass that would collapse in $\Delta t$
and create a collisionless particle:
$$\eqalignno{m_* = +m_b {\Delta t/t_{dyn}}  \hbox{\ \ \ and \ \ \ }\Delta m_b
&= -m_b {\Delta t/t_{dyn}} \qquad &(\the\eqtno) \cr}$$
\advance\eqtno by 1
at the center of the cell, giving it the same proper
velocity as the gas in the cell.
These collisionless particles are given three labels
at birth: their mass, $m_*$, the epoch of creation, $z_*$,
and $t_{dyn}$, the free-fall time in the birth cell.
After creation, these new particles are treated dynamically
the same as dark matter particles.
The three components (gas, galaxies and dark matter)
interact through gravity.
Feedback input into the IGM from young stars,
through the processes such as UV and supernovae,
is allowed for through
adoption of some fairly conservative values (see below)
of ``efficiencies" utilized
to parameterize these
processes.

We adopt conventional CDM parameters
$h=0.5$,
$\Omega=1$,
$\Omega_b=0.06$,
$\sigma_8=0.77$,
so the amplitude is somewhat ($35\%$) below
the COBE normalization but within
$2.5\sigma$ of the COBE results.
For the scales of interest the power spectrum adopted
is close to that in the COBE normalized
variants of CDM, namely the tilted, mixed,
and low $\Omega$ models all of which are adjusted
to give $\sigma_8\sim 0.8$ to match observational constraint.
The baryon density is taken from
standard light element nuelosythesis
(Walker \etal 1991).
Our box size is $80h^{-1}$\mpc, so,
with $200^3$ cells and dark matter particles,
our nominal
resolution is $400h^{-1}$kpc,
but actual resolution,
as determined by extensive tests (Cen 1992),
is approximately of a factor of $2.5$ worse than this ($\sim1h^{-1}$\mpc).
In addition, we have made a higher
resolution run with box size $8h^{-1}$\mpc,
the same number of cells and nominal
resolution $40h^{-1}$kpc.
In the small box
galaxy formation occurs earlier (because there are more nonlinear
waves in this box at earlier times)
and more vigorously
There is about 40\% less galaxy formation
in the $L=80h^{-1}$\mpc than
in the $L=8h^{-1}$\mpc,
partly due to the missing waves with wavelengths larger
than $L=8h^{-1}$\mpc which, if present as they should be,
would have heated the temperature to a higher level to reduce galaxy formation.
We ran the $L=8h^{-1}$\mpc box first,
then we added the UV/X-ray emissivities from sources
in the smaller box
to the larger box when the $L=80h^{-1}$\mpc simulation was run.
After the galaxy subunits are
created, they release energy in two forms:  UV from young stars and thermal
energy from
supernovae shocks (see Cen \& Ostriker, 1993b, for details).

For comparison with the
hydrodynamical simulations, we have run in parallel a
conventionally biased CDM N-body simulation with also
$\Omega = 1$, $h = 0.5$ and $b = 1.3$.  The CDM
evolves under the influence of gravity.
Some CDM particles are tagged in the initial
conditions (at z=20) as being peaks in
the initial conditions above a certain threshold where
the cell size of $0.4 h^{-1}$\mpc$/(1+z_{initial})$
is used with no smoothing.  In order to achieve the
desired bias of $b=1.3$,
the lower threshold for peaks to become galaxies is
$\rho > 0.92\bar\rho$
where $\bar\rho$ is the mean density at $z = 20$.
The biased particles are divided into groups.  The
top 25\% (elliptical galaxies) are those
peaks with $\rho > 1.735\bar\rho$,
the next 25\% (S0 galaxies)
are those peaks with $1.735\bar\rho >\rho >1.653\bar\rho$,
and the bottom 50\% (spiral galaxies) are those
peaks with $1.653\bar\rho  > \rho > 0.92\bar\rho$
where density $\rho$ and mean density
$\bar\rho$  in the universe are
both evaluated at z = 20.
Now, for a perturbation that is bound,
its radius $a_p(t)$ is given
parametrically by (cf. Gunn \& Gott 1972)
$$\eqalignno{a_p(\eta)&= {1\over 2} a_{max}(1-\cos \eta)\qquad &(\the\eqtno)
\cr}$$
\advance\eqtno by 1
$$\eqalignno{t_p(\eta)&= {1\over 2\pi} T_c(\eta-\sin \eta)\qquad, &(\the\eqtno)
\cr}$$
\advance\eqtno by 1
where $T_c$ is the collapse time
and
$a_{max}$ is the maximum radius, achieved when
$t = T_c/2$.
Meanwhile the mean expansion of the
$\Omega = 1$ universe is given by
$$\eqalignno{a(t)&= {a_{max}\over 4} ({12\pi t\over T_c})^{2/3}\qquad,
&(\the\eqtno) \cr}$$
\advance\eqtno by 1
\noindent which in the limit
of early times agrees with that of the perturbation.  Now
$$\eqalignno{\rho_p/\bar\rho&= a^3(t)/a_p^3(t)={9\over
2}{(\eta-\sin\eta)^2\over(1-\cos\eta)^3}\qquad, &(\the\eqtno) \cr}$$
\advance\eqtno by 1
\noindent Thus at maximum expansion at
$t = T_c/2$, $\eta = \pi$,
$a_p = a_{max}$,
$\rho_p/\bar\rho = 5.55$
(cf. Gunn \& Gott 1972),
and
$\rho_p/\bar\rho = 1.735$ at
$t = 0.1458 T_c$,
and
$\rho_p/\bar\rho  = 1.653$ at
$t = 0.1297 T_c$,
where
$t = t_o (1/21)^{3/2} = 0.135$ billion years
at $z = 20$ and
$t_o = 13$
billion years is the current
age of the universe in this
$\Omega = 1$, $h = 0.5$ model.
Thus, ellipticals are simply those galaxies
with collapse times
$T_c < 0.93$ billion years which form at
$z_c > 4.8$ (i.e., those peaks with
$\rho/ \bar\rho > 1.735$ at $z = 20$).
This is as in the Gott \& Thuan (1976)
proposal that ellipticals are
just those galaxies that have the shortest collapse times.
Then S0's are those galaxies that
have collapse times
$0.93$ billion years
$< T_c < 1.04$
billion years which form at
$4.38 < z_c < 4.8$
(i.e. those peaks with
$1.653 < \rho/\bar\rho < 1.735$ at $z = 20$).
Finally, spirals are those
galaxies that have collapse times that are longer than
$1.04$ billion years and form at
$z_c < 4.38$.
By way of comparison, the (25\%) ellipticals formed in the more realistic
hydrodynamic simulation formed in the interval
$7.5 > z_c > 3.6$, the (25\%) S0's formed in
the interval
$3.6 > z_c > 2.7$,
and the (50\%) spirals formed since
$z_c= 2.7$.
Thus the simple
biased N-body model with the biased particles divided into quartiles provides
an ad hoc
way of producing ellipticals, S0's and spirals which might naturally produce a
reasonable
morphology-density relation for these galaxies today but without following the
detailed
physics supplied by the hydrodynamic simulations.

	We will measure the topology of the large-scale structure outlined by
elliptical, S0,
and spiral galaxies, both as produced by the hydrodynamical simulations and by
the
simpler, more naive, biased N-body simulations.  This will show whether the
detailed
physics of galaxy formation on caustics and then galaxies draining onto
clusters by gravity
which occurs in the hydrodynamic simulations leaves a trace on the topology
which is
noticeably different from that produced by a simple N-body biasing scheme in
which
galaxy formation is dictated by collapse-time arguments alone.

\medskip
\centerline{3. TOPOLOGICAL TECHNIQUES}
\medskip

	Important clues as to the origin of large-scale structure
lie in its topology.  Gott, Melott, \& Dickinson (1986), Hamilton, Gott, \&
Weinberg (1986), and Gott et al. (1989)
have shown how topology can be measured using the genus statistic.  First, the
galaxy
distribution is smoothed with a Gaussian window function
$$\eqalignno{W(r)&= {1\over (2\pi)^{3/2}} \exp(-r^2/2\lambda^2)\qquad,
&(\the\eqtno) \cr}$$
\advance\eqtno by 1
\noindent where smoothing length
$\lambda$ is chosen so that
$2\lambda^2 = \lambda'^2 > l^2$
where the maximum mean
separation between galaxies in the sample is defined as
$l = n_{ex}^{-1/3}$
where
$n_{ex}$ is the
minimum expected volume density of galaxies anywhere in the sample (usually at
the outer
edge) calculated from the selection function.
For volume-limited samples with N galaxies
brighter than some limiting absolute magnitude within a volume
$V,l = (V/N)^{1/3}$.
Note:  we are using a new definition of the smoothing
length $\lambda$ as defined in equation (9), our old
smoothing length was
$\lambda'^2 = 2\lambda^2$,
so in previous papers such as Gott et al. 1989, when we
have used $\lambda' = 6 h^{-1}$\mpc,
this would correspond to
$\lambda = 4.24 h^{-1}$\mpc
with the new definition.
Since in most magnitude- or volume-limited samples
$l=5h^{-1}$\mpc,
we expect to use smoothing lengths
$\lambda > 3.5 h^{-1}$\mpc.
We use smoothing because we wish to study
large-scale structure and are not interested in the fact that the galaxies are
isolated points.
Once the smoothing has been done, density-contour surfaces in the smoothed
density field
are computed and are identified by a
variable
$\nu$ which is a measure of the volume fraction $f_v$
contained on the low-density side of the density contour surface:
$$\eqalignno{f_\nu&= {1\over \sqrt{2\pi}}\int_{-\infty}^\nu
\exp(-t^2/2)dt\qquad. &(\the\eqtno) \cr}$$
\advance\eqtno by 1
\noindent Thus
$\nu = 0$ is the median density contour which contains 50\% of the volume.
[This definition is such that if the density
field were gaussian random phase then
$\rho_{surface} - \bar\rho   = \nu (\delta \rho)_{rms}$].
The ($\nu = -2$,
$\nu = -1$, $\nu = 0$, $\nu = 1$ and $\nu = 2$)
surfaces contain respectively (2.5\%, 16\%, 50\%, 84\%,
and 97.5\%) of the volume on their low density
sides.
Alternatively, we may define the contours by the fraction of mass
$f_M$ of the smoothed-galaxy mass distribution contained
on their low-density sides.

	The topology of each density-contour surface is measured by the genus
statistic.
$$\eqalignno{G&= \hbox{Number of ``donut" holes} - \hbox{Number of isolated
regions.} &(\the\eqtno) \cr}$$
\advance\eqtno by 1
\noindent By this definition, a sphere with $N$ handles would have a genus of
$G = N-1$ because it is
one isolated region and has N holes.
Also by this same definition, $3$ spheres have a genus
of $G = -3$
because they have no holes and are 3 isolated regions.  As shown by Gott,
Melott, \& Dickinson (1986) using the Gauss-Bonnet Theorem
$$\eqalignno{G&= -{1\over 4\pi} \int KdA\qquad, &(\the\eqtno) \cr}$$
\advance\eqtno by 1
\noindent where
$K = 1/r_1r_2$
is the Gaussian curvature, and
$r_1$ and $r_2$ are the two principle radii of
curvature, and the integral is carried out over the entire surface area of the
contour surface.
Thus 1 sphere has a genus
$G = -1$ because $K = 1/r^2$,
$\int dA = 4\pi r^2$,
so by eq. 12
$G = -1$.
The above formula allows us to develop a computer
program which goes pixel by pixel in a
smoothed density array and computes the genus for each density-contour surface
(see Gott,
Melott, \& Dickinson 1986, and Weinberg 1988 for a copy of the program called
CONTOUR).

	What if we did such a topology study in the initial conditions in a standard
inflationary model where the fluctuations
are produced by quantum noise and are gaussian
random phase?  Then we expect that the
average genus per unit volume
$g(\nu)$ is given by
$$\eqalignno{g(\nu)&=A(1-\nu^2)\exp(-\nu^2/2)\qquad, &(\the\eqtno) \cr}$$
\advance\eqtno by 1
\noindent where
$$\eqalignno{A&={1\over 4\pi^2}\bigl({1\over 3}{\int k^2P'(k)d^3k\over \int
P'(k) d^3k}\bigr)^{3/2}\qquad, &(\the\eqtno) \cr}$$
\advance\eqtno by 1
\noindent and where
$P'(k) = P(k) \exp (-k^2\lambda^2)$
is the smoothed power spectrum and
$P(k)$ is the primordial power spectrum
(Hamilton, Gott, \& Weinberg 1986).
We note that while the
amplitude of the genus curve A depends on the power spectrum
[for
$P(k) \propto k^n$,
$A = ([3+n]/6)^{3/2}/4\pi^2 \lambda^3]$
- with a smaller amplitude for models with more power at large scales -
the form of the genus curve as a function of
$\nu$ is independent of the power spectrum
$g(\nu) \propto (1-\nu^2) \exp (-\nu^2/2)$.
It is positive for $-1 < \nu <1$,
indicating holes and a spongelike
topology, it is negative for
$\nu > 1$
indicating isolated clusters, and negative for $\nu < 1$
indicating isolated voids.
Now, while the fluctuations are in the linear regime, they simply
grow in place increasing in amplitude
$(\delta\rho/\rho \propto a)$ so that the topology does not change at all.
If $r_o$ is the correlation length at the present epoch,
then $\xi(r_o)=1$, and if we choose
$\sqrt{2}\lambda > r_o$,
we can guarantee that the fluctuations we are looking at are still
approximately in the linear
regime.  To lower statistical noise the curve is smoothed using boxcar
averaging
$g(\nu) = 1/3 [g (\nu-0.1) + g(\nu) + g(\nu+0.1)]$
 (cf. Vogeley et al. 1994).

	Since making the prediction that the topology of large-scale structure should
be
spongelike and approximately random phase (Gott, Melott, \& Dickinson 1986;
Hamilton,
Gott, \& Weinberg 1986; Gott, Weinberg, \& Melott 1987), this prediction has
been
confirmed in every observational sample studied:
[CfA, Giovanelli \& Haynes, Tully,
Thuan \& Schneider, and Abell cluster samples] Gott et al. (1989);
[IRAS] Moore et al. (1992);
[Abell clusters] Rhoads, Gott, \& Postman (1994);
[CfA 1+2] Vogeley et al. (1994).
In all cases,
$g(\nu=0) > 0$,
indicating a spongelike topology for the median density contour.

	We examine the galaxies produced in the hydrodynamical simulations as they
appear at the present epoch.  This simulation has
$\Omega_{CDM} = 1$,
$h = 0.5$
and amplitude $\sigma_8 =1/b = 0.77$
(i.e., $(\delta\rho/\rho)_{CDM} = 1/1.3$
for spheres of radius
$8 h^{-1}$ \mpc).
The galaxies that have formed and evolved in the
hydrodynamical simulations are observed to have
$(\delta\rho/\rho)_{gal} = 1$
for spheres of radius
$8 h^{-1}$\mpc, giving a natural bias factor
of $1.3$.
We regard this particular model, not as necessarily the best or only one, but
following the discussion in the introduction, to simply be representative of
the physics
obtained in the broad class of standard inflationary CDM models (with
$\Omega$ not necessarily 1)
in which the structure originates from random quantum noise in the early
universe.
Starting with a
magnitude-limited survey with an optimally chosen outer boundary, one can
typically
produce a volume-limited survey with a mean separation of bright galaxies of
$l = 5 h^{-1}$\mpc.
Thus our computational volume is simulating a volume-limited observational
sample of
4096 galaxies (which could be drawn from a magnitude-limited survey of 16,000
galaxies)
and is just somewhat larger in volume than the CfA 1+2 sample.
We divide this sample, as noted,
into quartiles by epoch of formation:  the earliest forming 25\% we will call
ellipticals, the
second 25\%, S0's, and the final 50\%, spirals.
Figure 1 shows the distribution of these subsamples of  galaxies
in a slice of size $80\times 80\times 15h^{-3}$\mpc$^3$
  Now for each quartile the mean separation
between galaxies  within that quartile is
$l = (4)^{1/3} 5 h^{-1}$\mpc$= 7.9 h^{-1}$\mpc.  To avoid
discreteness effects we therefore should adopt
$\sqrt{2}\lambda > l$,
$\lambda > 5.6 h^{-1}$ \mpc.  To be on the
cautious side we will adopt
$\lambda = 8 h^{-1}$ \mpc.  This is large enough that we may ignore
peculiar velocities and, since
$\lambda > r_o$,
we are also assured that we are looking at scales
where the fluctuations are still approximately in the linear regime and we may
expect the
random-phase topology formula to approximately hold true except for small
deviations
caused by non-linear effects (which are precisely those we are investigating).

\medskip
\centerline{4.  RESULTS}
\medskip

	Figure 2 (solid line)
 shows the genus curve for the CDM in the sample smoothed at
$\lambda = 8 h^{-1}\mpc = 800 \kms$.
It is approximately random-phase - the best fitting theoretical random
phase genus curve
$[g(\nu) \propto (1-\nu^2) \exp (-\nu^2/2)]$
is shown for comparison (dashed line).

	Figure 3 shows the genus curve (and best fit theoretical random phase curve)
for all
galaxies in the hydrodynamic simulations.  In these genus curves and all that
follow, the
total genus for the
$(80 h^{-1} \mpc)^3$
volume is recorded.  The error bars are computed by
dividing the cube into 8 subcubes, each of
$(40 h^{-1} \mpc)^3$
volume, and computing the
standard deviation of the mean genus per unit volume for the entire cube from
these 8
independent estimates.  The smoothing length is
$\lambda = 8 h^{-1}\mpc = 800 \kms$ in all cases.
Both the CDM and the galaxies approximately
follow the random-phase curve.
It is significant
that the galaxies still approximate the random-phase curve,
for it shows that the non-linear
hydrodynamic effects in the simulations do not wipe out the influence of the
random phase
initial conditions.  This means that we can still use the genus curve
as a test of whether or
not the initial conditions were random phase.  If larger smoothing lengths were
used, the
agreement with the theoretical curve would be better and better, provided that
large enough
survey volumes were used.

	Figures 4a,b,c show the genus curves
of each group of galaxies
classified by epoch of
formation:  4a, oldest 25\% = ellipticals; 4b, second 25\% = S0's; and 4c, last
50\% =
spirals.  The spirals are approximately random phase in reasonable agreement
with the
expected spongelike topology.

	The ellipticals show a significant
systematic shift toward the left called a "meatball shift" (cf.
Weinberg, Gott \& Melott 1987; Gott et al. 1989).
If galaxies are placed down in a meatball
distribution these isolated clusters will retain their identity (and provide
negative genus
contour surfaces, indicating isolated high-density regions) down to a lower
threshold value
of
$\nu$ than would otherwise be the case.  In a
random-phase distribution with
$g(\nu) \propto (1 - \nu^2) \exp (-\nu^2/2)$
we expect to find
$g(\nu) < 0$
indicating isolated clusters for $\nu > 1$.
For a meatball
topology we expect
$g(\nu) < 0$ for
$\nu > \nu_+$
where $\nu_+ < 1$ (cf. Weinberg, Gott, \& Melott 1987).
This has the effect of pushing the entire genus curve to the left, in
particular, moving the
peak of the genus-curve to
$\nu < 0$
and compressing the region of negative genus isolated
voids
$g(\nu) < 0$,
for $\nu < \nu_-$ where
$\nu_- < - 1$.
This shows that the older
galaxies designated as ``ellipticals"
in the
hydrodynamic simulations are preferentially located
in clusters relative to the younger galaxies designated
as ``spirals" which
have a more nearly random-phase spongelike topology characteristic of a
filamentary net.
The ``S0's" are intermediate.

	We can quantify this shift by use of the shift statistic
$\Delta \nu$ (cf. Vogeley et al, 1994)
tabulated in Table 1.
$$\eqalignno{\Delta \nu&=\int_{\nu=-1}^{\nu=+1}\nu g(\nu)
d\nu/\int_{\nu=-1}^{\nu=+1}g_{fit}(\nu) d\nu\qquad. &(\the\eqtno) \cr}$$
\advance\eqtno by 1
This gives the mean shift in the genus curve.
A negative value of $\Delta\nu$ indicates a shift to the
left or a meatball shift, a positive value of
$\Delta\nu$ indicates a shift to right or a ``swiss cheese"
shift (i.e., toward a swiss cheese topology of isolated voids - see Gott,
Weinberg, \&
Melott 1987).  For ellipticals
$\Delta\nu = - 0.150$, while for spirals
$\Delta\nu = - 0.0864$, showing that
the ellipticals have a larger meatball shift than the spirals, apparent from a
visual inspection
of the genus curves in fig. 4a and 4c.

	To further illustrate, in fig. 5a and b we show the
$\nu = - 1$ density contour surfaces
for the ellipticals and the spirals.  This contour contains 16\% of the volume
on its low-
density side and in a random-phase distribution would mark the transition from
isolated
voids to a spongelike topology.  Because of the meatball shift of the genus
curve of the
ellipticals relative to the spirals, this particular contour, for the
ellipticals, is more multiply
connected (has more ``donut" holes) than the surface for spiral galaxies.  This
is shown by
the tube connecting the two voids at the bottom and on the right of the
elliptical cube (which
adds +1 to the genus) which is absent in the spiral cube, where those two voids
are
separate.

	Other statistics in table 1 include the best fit amplitude of the random phase
curve as
a fraction of the amplitude expected from the initial power spectrum (i.e., the
amplitude that
would have been measured for the initial conditions themselves).  A value of
$R_G < 1$
indicates there has been some non-linear merging of structures, a little of
which is to be
expected (Melott, Weinberg, \& Gott 1988).  Finally, there is the width of the
genus curve
$W_\nu = \nu_2 - \nu_1$
which  measures the width of the region that has positive genus or spongelike
topology.  For a random-phase distribution
$g(\nu) > 0$ for
$- 1 < \nu < 1$
so
$W_\nu = 2.0$.
If $W_\nu > 2.0$, as is true for these distributions, it indicates a
filimentary net which retains its
spongelike topology over a larger range of
$\nu$ than would be true in a random phase
distribution.  The CfA 1 + 2 sample for all galaxies also shows a
$W_\nu$ value somewhat
larger than 2 (Vogeley et al. 1994).

	Another way to illustrate the difference in the distribution between
ellipticals and
spirals in the hydrodynamical simulations is to note that each density-contour
surface can
be labeled either by the volume fraction
$f_V$ contained on its low-density side or by the mass
fraction $f_M$ contained on its low-density side.  Each density contour surface
thus appears as
a point on the
$f_V$,
$f_M$
plane and the entire family of density contours for (ellipticals, say, or
spirals) can be plotted as a curve in the $f_V$,
$f_M$ plane (see fig. 6).  In the initial conditions
$f_V \approx f_M$, so departures from this line are entirely due to non-linear
effects.  The curves for
dark matter, all galaxies, ellipticals, and spirals are shown.  The curve for
ellipticals
deviates more from the $f_V = f_M$ line than any of the others, showing them to
be more
strongly clumped.  This is in line with the morphology-density relation shown
by the
hydrodynamic simulation which is in agreement with that observed.

	The difference in topology in the hydrodynamical simulations between the
ellipticals
and the spirals can be displayed most graphically by plotting genus versus
$f_M$ for each as is
done in figures 7a and b.  The ellipticals show a positive genus (spongelike
topology) for
$0.03 < f_M < 0.58$; the genus is negative for
$f_M > 0.58$.  This means that 42\% of the
smoothed mass distribution in ellipticals is contained in isolated clusters.
For spirals, the
genus is positive for
$0.06 < f_M < 0.70$; the genus is negative for
$f_M > 0.70$.  This means
that only 30\% of the smoothed mass distribution in spirals is contained in
isolated clusters.
This quantifies the propensity for the ellipticals to be found in clusters.

	The genus curves for ellipticals and spirals in the biased N-body simulations
are
shown in figures 8a and c.  There is not any dramatic meatball shift of the
ellipticals relative
to spirals.  This is quantified in Table 1:
$\Delta\nu = - 0.0455$ for ellipticals and $\Delta\nu = - 0.0671 $
for spirals.  By comparison, in the hydrodynamic simulations,
$\Delta\nu = - 0.150$ for ellipticals
and
$\Delta\nu = - 0.0864$ for spirals, giving a much larger differential
difference.

	Genus versus $f_M$ curves drawn from the biased N-body simulation (shown in
figs.
9a and b) are more nearly similar to each other and do not show such a large
difference as
seen in the hydronamic simulations.  For these ellipticals, the genus is
negative for
$f_M > 0.68$, and for these spirals, the genus is negative for
$f_M > 0.71$.
Thus 32\% of the
smoothed density distribution of ellipticals is in isolated clusters versus
29\% for spirals - a
difference of only 3\%  as opposed to the 12\% difference predicted by the
hydrodynamic
simulations.  Thus we could test future observational samples for this effect
by plotting the
genus as a function of
$f_M$ for both ellipticals and spirals, and then comparing them.

	Topology can of course be studied as a function of
smoothing length $\lambda$
(or $\lambda' = (2)^{1/2}\lambda$).
For $\lambda > 800 \kms$
the elliptical and spiral curves should both become more
nearly random phase and therefore more nearly alike.  As studies we have done
by plotting
$f_M$ versus $f_v$ for various smoothing lengths indicate, for
$\lambda < 800 \kms$
the elliptical and
spiral curves become more and more differentiated.  The smallest smoothing
length one
would want to use for a topology study by galaxy type would be
$\lambda' = l_e = (4)^{1/3} 5h^{-1}$\mpc,
or $\lambda = 560 \kms$,
where
$l_e$ is the minimum mean distance between ellipticals in a
typical volume or magnitude limited
sample.  Smaller than this we would begin to
encounter discreteness effects.  At a smoothing length of
$\lambda = 560 \kms$,
some care would
also have to be taken to ensure that the simulation had a peculiar velocity
dispersion for
ellipticals that was in agreement with the observed sample, since the line of
sight rms
peculiar velocity dispersion for ellipticals in the simulations (or
observations) might be as
large as
$560 \kms$.
Also, operating at the edge where discreteness effects might become
important at
$\lambda = 560 \kms$
would require more accurate simulation of the luminosity
function.  As a further check against discreteness effects, we could divide the
spirals into 2
halves and check that the ellipticals show the same meatball shift relative to
each half of the
spirals and that there is no difference between the average of the genus curves
for the spiral
halves, and that for all the spirals taken together.  (This is certainly the
case at
$\lambda = 800 \kms$.)
Against these cautions would be the benefits of having a larger mean difference
expected between the elliptical and spiral curves
and smaller error bars.  Since the errors are
primarily counting errors and since for a power spectrum
$P(k) \sim k^{-1}$ (for CDM at these
scales) we expect
$A\propto V/\lambda^3$,
and so $\delta g/g \propto V^{1/2} \lambda^{-3/2}$.
So going from
$\lambda = 800 \kms$
to $\lambda = 560 \kms$
could lower the relative statistical errors by a factor of 1.7.
Figures (10a,b) show the equivalent of Figures (7a,b)
but with $\lambda = 560 \kms$.
As expected,
the elliptical and spiral genus curves are even
more differentiated from each other.
At this smoothing length $52\%$ of the smoothed mass distribution
in the ellipticals is in isolated clusters whereas only
$37\%$ of the smoothed mass distribution
in the spirals is in isolated clusters (a difference of $15\%$).
Alternatively, as we shall discuss
below, the simplest way to lower the relative statistical errors is simply to
study a larger
survey volume - which should be possible in the near future.

\medskip
\centerline{5.  CONCLUSIONS}
\medskip

	Topology can provide many important clues as to the formation of large-scale
structure in the universe.  Hydrodynamic simulations with CDM and baryons,
following
known physics, give rise to structures quite promisingly like those seen in the
observations.  Starting with Gaussian
random-phase initial conditions, fluctuations
initially grow entirely
by gravitational instability, caustics and shocks
form when the fluctuations go non-linear.
Galaxies are formed in these shocks and
so are initially located along walls and filaments.
This picture developed in Cen \& Ostriker (1993a) and
supported by the present work is the following.
The first-forming galaxies are identified
with ellipticals.  Later, S0's and finally spirals
form.
This designation in the simulation produces samples which
qualitatively reproduce the
properties of real ellipticals, S0's and spirals
with regard to the density-morphology relation,
gas-to-total mass ratio, dark matter-to-stellar ratio and
other properties.
  After formation, galaxies, under the action of gravity, drain off walls onto
filaments
and then into clusters.  Ellipticals, which get a head start on this process,
are therefore seen
at the present epoch in the simulations preferentially in clusters relative to
spirals which are
still primarily in a filamentary net.  The topology can be measured
quantitatively with the
genus statistic.  Random-phase initial conditions produce a characteristic
spongelike
topology if the density field is smoothed on a scale larger than the mean
intergalactic
distance and larger than the correlation length.  That is because in the
gravitational
instability picture, fluctuations grow in place (increasing in amplitude) as
long as they are
still in the linear regime and they retain the random-phase topology they
inherited from the
initial conditions.  If genus is measured as a function of density, we expect
to find isolated
voids at very low density, isolated clusters at very high density, and we
expect the median
density-contour surface to be spongelike.
This produces a symmetric genus curve
$g(\nu) \propto (1 - \nu^2) \exp (- \nu^2/2)$.
The hydrodynamic simulations show for all galaxies a genus curve
that approximately follows this random phase law if a smoothing length of
$\lambda = 8 h^{-1}$\mpc
is adopted.  This shows that the hydrodynamic effects
do not mask the topology inherited
from the initial conditions and that measuring the genus curve can be used as a
test for
whether the initial conditions were random phase or not.  This is quite
important since
inflationary models in which structure arises from random quantum functions in
the early
universe do have random-phase initial conditions wheras
certain other models (e.g., textures, cosmic strings, domain walls) do not.
The random-phase nature of the initial fluctuations can be
checked independently by using a 2D genus statistic (hot spots - cold spots)
(Gott et al.
1990) on the fluctuations seen in the cosmic microwave background.  Three
groups (cf.
Torres 1994; Smoot et al 1994; Park \& Gott 1994) have independently shown that
the year-
1 COBE microwave background sky maps give an excellent fit to the theoretical
random-
phase curve (Melott et al. 1989) predicted for this statistic
$g_{2D}(\nu) = \nu \exp (-\nu^2/2)$.

	As expected, however, the simulations do show some small but measurable
deviations from the random-phase
theoretical genus curve due to non-linear effects.
The genus curve for ``ellipticals" is shifted to the left
$(\Delta\nu = - 0.150)$ more than that for the ``spirals"
$(\Delta\nu = - 0.086)$.
This shows the greater propensity for the ellipticals to be found in
clusters.  The difference between the topology
of the ellipticals and the spirals shows up
most dramatically when we measure the genus of contour surfaces
as a function of the
mass fraction in the smoothed distribution.
This treatment shows that 42\% of the
smoothed mass distribution of ellipticals is in
isolated clusters as opposed to only 30\% of
spirals.
These topological properties are signs
of the physical origin of the
``bias" found in the hydrodynamic simulations.
Effects this large do not show up in standard biased N-body simulations
based on a peaks approach.

Studies of the topology
of large-scale structure by galaxy type in future observational
samples should easily be able to test for this effect.
If, as expected, real ellipticals show a meatball shift similar
to the ``elliptical" subset in the simulation,
and dissimilar to the ``elliptical" set
picked out by standard peak biasing technique,
this would provide evidence for the
physical plausibility of the hydrodynamic simulations and for the
designation of ellipticals as a
statistically old population of galaxies.
The current CfA 1+2 sample is almost
as large in volume as these simulations and the Sloan Digital Sky Survey (SDSS)
will be
62 times larger in volume.
 Automated typing algorithms will be implemented to type the
galaxies in the SDSS from the imaging phase of the survey.  This should allow
us to
implement topology studies by galaxy type as described in this paper on a grand
scale.
Likewise, over the next few years computer hydrodynamic simulations are likely
to also
improve by a large factor from $(200)^3$
grids to
$\sim (1600)^3$ grids allowing more resolution
($200 h^{-1}$kpc)
while giving survey volumes
$64$ times larger than at present.
New effects
may also be added such as ram pressure stripping of
spirals falling into clusters (cf. Gunn
\& Gott 1972), converting them to S0's.

	Currently observational samples show
approximately random-phase genus curves
(Gott et al. 1989; Moore et al. 1992; Vogeley et al. 1994).  This supports the
claim that the
structures we see today originated from small-amplitude Gaussian random initial
conditions.  Although all observational samples are approximately random phase
(all have a
spongelike median-density-contour surface, for example), some small differences
can be
noted.  In general, samples including all galaxies are either random phase or
show a small
shift in the direction of a meatball topology
$(\Delta\nu < 0)$ like that seen in the ellipticals and
spirals in the hydrodynamic simulations.
The largest sample, the CfA 1+2, has a genus
curve with a width $W_\nu$ - somewhat wider
than that for a random-phase distribution, an
effect also seen in these hydrodynamic simulations.
Since the elliptical curve is shifted
relative to the spiral curve, when both populations
are combined, an overall wider genus
curve should be produced.  The amplitude of the genus
curve for the CfA 1+2 is best fit by
a CDM model with
$\Omega h = 0.3$.
Such models, with more power at large scales than the
hydrodynamic
$\Omega h = 0.5$ CDM model,
are expected to show for all galaxies a genus curve
that is less meatball shifted overall (cf. Vogeley et al. 1994).  We would
expect the same
difference between ellipticals and spirals, but for an
$\Omega h = 0.3$ model both populations
would show somewhat less propensity to be found in isolated clusters.  So if
the galaxies
were typed, we might expect that,
 with $\lambda = 8 h^{-1}$\mpc smoothing, somewhat less than
42\% of the smoothed mass fraction in ellipticals would be in isolated clusters
and
somewhat less than 30\% of the smoothed mass fraction in spirals would be in
isolated
clusters, but an overall difference of 12\% in the fraction of each population
in isolated
clusters might still be expected.  In the future, comparison of newer, larger,
even more
realistic hydrodynamic simulations with much larger observational samples using
these
topological techniques - each
surveying a volume $\sim 64$ times as large as that depicted here
(and with error bars on all graphs 8 times as small
relative to the curves) offers exciting
possibilities for testing
and illuminating the details of the galaxy formation process.

This work is supported in part by
NASA grant NAGW-765, NAGW-2448 and NAG5-2759,
NSF grants AST90-20506, AST91-08103 and ASC-9318185,
and W. M. Keck Foundation.
We would like to thank Kevin Perry
for his assistance with the color graphics.
RC and JPO would like to thank the hospitality of ITP during their stay
when this work was completed, and the financial support from ITP
through the NSF grant PHY94-07194.
\vfill\eject

\centerline{REFERENCES}
\medskip
\refset
Bahcall, N.A., \& Cen, R. 1992, ApJ, 398, L81
\medskip
\refset
Bardeen, J.M., Bond, J.R., Kaiser, N., \& Szalay, A.S. 1986, ApJ, 304, 15.
\smallskip
\refset
Bardeen, J.M., Steinhardt, P.J., \& Turner, M.S. 1983, Phys. Rev. D., 28, 679.
\smallskip
\refset
Bartlett, J.G., Blanchard, A., Silk, J., \& Turner, M. 1995, Science, 267, 980.
\smallskip
\refset
Bhavsar, S.P. 1981, ApJ, 246, L5.
\smallskip
\refset
Broadhurst, T.J., Ellis, R.S., Koo, D.C., \& Szalay, A.S. 1990, Nature, 343,
726.
\smallskip
\refset
Bucher, M., Goldhaber, A.S., Turok, N., submitted to Physical Rev. D.
\smallskip
\refset
Cen, R. 1992, ApJS, 78, 341.
\smallskip
\refset
Cen, R., \& Ostriker, J.P. 1992a, ApJ 393, 22.
\smallskip
\refset
Cen, R., \& Ostriker, J.P. 1992b, ApJ, 399, L113.
\smallskip
\refset
Cen, R., \& Ostriker, J.P. 1993a, ApJ, 417, 415.
\smallskip
\refset
Cen, R., \& Ostriker, J.P. 1993b, ApJ, 417, 404.
\smallskip
\refset
Cen, R., \& Ostriker, J.P. 1994, ApJ, 431, 451.
\smallskip
\refset
Cen, R., Gnedin, N., \& Ostriker, J.P. 1993 ApJ, 417, 387.
\smallskip
\refset
Cen, R., Miralda-Escude, J., Ostriker, J.P, \& Rauch, M. 1994, ApJ, 437, L2.
\smallskip
\refset
Cen, R., Kang, H., Ostriker, J.P, \& Ryu, D. 1995, ApJ, submitted
\smallskip
\refset
Davis, M., \& Geller, M.J. 1976, ApJ, 208, 13.
\smallskip
\refset
de Lapparent, V., Geller, M.J., \& Huchra, J.P. 1986, ApJ, 302, L1.
\smallskip
\refset
Doroshkevich, A.G., Zeldovich, Ya.B., \& Sunyaev, R.A. 1976, Formation and
Evolution
of Galaxies and Stars, Moscow.
\smallskip
\refset
Dressler, A. 1980 , ApJ, 236, 351.
\smallskip
\refset
Eggen, O.J., Lynden-Bell, D., \& Sandage A.R. 1962, ApJ, 136, 748.
\smallskip
\refset
Efstathiou, G., Bond, J.R., \& White, S.D.M. 1992, MNRAS, 258, 1p
\smallskip
\refset
Einasto, J., Joeveer, M., \& Saar, E. 1980, MNRAS, 193, 353.
\smallskip
\refset
Freedman, W.L. et al. 1994, Nature, 371, 757.
\smallskip
\refset
Geller, M.J., \& Huchra, J.P. 1989, Science, 246, 897.
\smallskip
\refset
Gott, J.R. 1977, Ann. Rev. As. Astrophysics, 15, 235.
\smallskip
\refset
Gott, J.R. 1980 Physical Cosmology, Les Houches, eds. J. Audouze, R. Balain, \&
D.N. Schramm (North Holland Publishing Company, Amsterdam) 561.
\smallskip
\refset
Gott, J.R. 1982 Nature, 295, 304.
\smallskip
\refset
Gott, J.R., \& Gunn, J.E. 1971ApJ, 169, L13.
\smallskip
\refset
Gott, J.R., Turner, E.L., \& Aarseth, S.J. 1979 ApJ, 234, 13.
\smallskip
\refset
Gott, J.R., Weinberg, D.H., \& Melott, A. 1987 ApJ, 319, 1.
\smallskip
\refset
Gott, J.R., \& Statler, T.S. 1984 Physics Letters, 136B, 157.
\smallskip
\refset
Gott, J.R., Miller, J., Thuan, T.X., Schneider, D.E., Weinberg, D.H.,
Gammie, C., Polk, K., Vogeley, M.,
Jeffrey, S., Bhavsar, S., Melott, A.L., Giovanelli, R.,
Haynes, M.P., Tully, R.B., \& Hamilton, A.J.S. 1989 ApJ, 340, 625.
\smallskip
\refset
Gott, J.R., Park, C., Juskiewicz, R., Bies, W.E., Bennett, D.P., Bouchet, F.R.,
\&
Stebbins, A. 1990, ApJ, 352, 1.
\smallskip
\refset
Gott, J.R. 1986 Inner Space/Outer Space, The Interface Between Cosmology and
Particle
Physics, eds. E.W. Kolb, M.S. Turner, D. Lindley, K. Olive, \& D. Seckel
(Chicago
and London, University of Chicago Press), 362.
\smallskip
\refset
Gott, J.R., Melott, A., \& Dickinson, M. 1986, ApJ, 306, 341.
\smallskip
\refset
Gott, J.R., \& Thuan, T.X. 1976, ApJ, 204, 649.
\smallskip
\refset
Groth, E.J., \& Peebles, P.J.E. 1977, ApJ, 217, 385.
\smallskip
\refset
Gunn, J.E., \&  Gott, J.R  1972, ApJ, 176, 1.
\smallskip
\refset
Hamilton, A.J.S., Gott, J.R., \& Weinberg, D. 1986, ApJ, 309, 1.
\smallskip
\refset
Jameson, A. 1989, Science, 245, 361.
\smallskip
\refset
Kamionkowski, M. \& Spergel, D. 1994, ApJ, 432, 1.
\smallskip
\refset
Kang, H., Ostriker, J.P., Cen, R., Ryu, D., Hernquist, L., Evard, A.E., Bryan,
G.L., \&
Norman, M.L. 1994, ApJ, 430, 83.
\smallskip
\refset
Klypin, A. Holtzman, J., Primack, J., \& Regos, E. 1993, ApJ, 416, 1.
\smallskip
\refset
Kofman, L., Gnedin, N.Y., \& Bahcall, N.A. 1993, ApJ, 413, 1
\smallskip
\refset
Ma, C.P., \& Bertschinger, E. 1994, ApJ, 435, L5
\smallskip
\refset
Maddox, S.J., Efstathiou, G., Sutherland, W.J.,
\& Loveday, J. 1990, MNRAS, 242, 43p.
\smallskip
\refset
Mather, J.C. et al. 1990, ApJ, 354, L37.
\smallskip
\refset
Melott, A.L., Weinberg, D.H., \& Gott, J.R. 1988, ApJ, 328, 50.
\smallskip
\refset
Melott, A.L., Cohen, A.P., Hamilton, A.J.S.,
Gott, J.R., \& Weinberg, D.H. 1989, ApJ, 345, 618.
\smallskip
\refset
Moore, B., Frenk, C.S., Weinberg, D.H., Saunders, W.,
Lawrence, A., Ellis, R.S., Kaiser, N., Efstathiou, G.,
\& Rowan-Robinson, M. 1992, MNRAS, 256, 477.
\smallskip
\refset
Ostriker, J.P. 1993, ARAA, 31, 689
\smallskip
\refset
Park, C. 1990, MNRAS, 242, 59P.
\smallskip
\refset
Park, C., \& Gott, J.R  1991, MNRAS, 249, 288.
\smallskip
\refset
Park, C., Gott, J.R., Melott, A., \& Karachentsev, I.D. 1992a, ApJ, 387, 1.
\smallskip
\refset
Park, C., Gott, J.R., da Costa, L.N. 1992b, ApJ, 392, L51.
\smallskip
\refset
Park, C., \& Gott, J.R. (unpublished).
\smallskip
\refset
Peebles, P.J.E. 1980, The Large-Scale Structure of the Universe, Princeton
University
Press, Princeton, NJ.
\smallskip
\refset
Peebles, P.J.E. 1982, ApJ, 263, L1.
\smallskip
\refset
Peebles, P.J.E. 1984, ApJ, 284, 439.
\smallskip
\refset
Pierce, M.J., Welch, D.L., McClure, R.D., van den Bergh, S., Racine, R., \&
Stetson,
P.B. 1994, Nature, 371, 385.
\smallskip
\refset
Ratra, B., \& Peebles, P.J.E. 1994, ApJ, 432, L5.
\smallskip
\refset
Rees, M.J., \& Ostriker 1977, MNRAS, 179, 451.
\smallskip
\refset
Rhoads, J., Gott, J.R., \& Postman, M. 1994, ApJ, 421, 1.
\smallskip
\refset
Sachs, R.K., \& Wolfe, A.M. 1967, ApJ, 147, 73.
\smallskip
\refset
Saunders, W., Frenk, C., Rowan-Robinson, M., Efstathiou, G., Lawrence, A.,
Kaiser,
N. Ellis, R., Crawford, J., Xia, X-Y, \& Parry, I. 1991, Nature, 349, 32.
\smallskip
\refset
Shectman, S.A., Landy, S.D., Oemler, A.A., Tucker, D., Kirshner, R.P., Lin, H.,
\&
Schechter, P.L. 1995, Harvard-Smithsonian Center for Astrophysics Preprint, to
appear in the 35th Herstmonceaux Conference, Wide Field Spectroscopy and the
Distant Universe.
\smallskip
\refset
Smoot, G.F. et al. 1992, ApJ, 396, L1.
\smallskip
\refset
Smoot, G.F., Tenorio, L., Banday, A.J., Kogut, A., Wright, E.L., Hinshaw, G.,
Bennett, C.L. 1994, ApJ, 437, 1.
\smallskip
\refset
Taylor, A.N., \& Rowan-Robinson, M. 1992, Nature, 359, 396.
\smallskip
\refset
Torres, S. 1994, ApJ, 423, L9.
\smallskip
\refset
Vogeley, M.S., Park, C., Geller, M.J., Huchra, J.P., \& Gott, J.R. 1994, ApJ,
420,
525.
\smallskip
\refset
Walker, T.P., Steigman, G., Schramm, D.N., Olive, K.A., \& Kang, H.S. 1991,
ApJ,
376, 51.
\smallskip
\refset
Weinberg, D. 1990, PhD Thesis, Princeton University.
\smallskip
\refset
Weinberg, D.H., \& Gunn, J.E. 1990, ApJ, 352, L25.
\smallskip
\refset
Weinberg, D.H. 1988, P.A.S.P., 100, 1373.
\smallskip
\refset
Weinberg, D.H., Gott, J.R., \& Melott, A.L. 1987, ApJ, 321, 2.
\smallskip
\refset
White, S.D.M., Frenk, C.S., Davis, M., \& Efstathiou, G. 1987, ApJ, 313, 505.
\smallskip
\refset
Yamamoto, K., Sasaki, M., \& Tanaka, T. 1995, preprint.
\smallskip
\refset
\vfill\eject

\hoffset=0.0truecm
\hsize=6.50truein
\centerline {{\bf Table 1.} Genus Curve Statistics}
\medskip
$$\vbox{\settabs 4 \columns
{\bigskip}
{\hrule height 0.4pt}
{\smallskip}
{\hrule height 0.4pt}
{\bigskip}
\+ & \hfill \vbox{\hbox{\hskip 1.cm Amplitude} \hbox{\hskip 1.cm \ \ \ \
$R_G$}}\hfill  & \hfill \vbox{\hbox{Width}\hbox{\ \ $W_\nu$}}\hfill  & \hfill
\vbox{\hbox{Shift}\hbox{\ $\Delta \nu$}} \hfill \cr
{\bigskip}
{\hrule height 0.4pt}
{\bigskip}
\+\hfill I. Hydrodynamic Simulation \hfill  & & & \cr
{\medskip}
\+ A. Cold Dark Matter\hfill  & \hfill\hskip 1.cm 0.844\hfill  & \hfill
2.06\hfill  & \hfill -0.0529\hfill  \cr
\+ B. All Galaxies \hfill & \hfill\hskip 1.cm  0.877\hfill  & \hfill 2.03\hfill
 & \hfill -0.119\hfill  \cr
\+ C. ``Ellipticals" (oldest 25\%)\hfill  & \hfill\hskip 1.cm  1.00\hfill  &
\hfill 2.03\hfill  & \hfill -0.150 \hfill \cr
\+ D. ``Spirals" (youngest 50\%)\hfill  &\hfill\hskip 1.cm   0.861\hfill  &
\hfill 1.978\hfill  & \hfill -0.0864\hfill  \cr
{\bigskip}
{\bigskip}
{\bigskip}
\+\hfill II.  Biased N-Body Simulation\hfill  & & & \cr
{\medskip}
\+ All Initial Bias Particles (IBP)\hfill &\hfill\hskip 1.cm 1.189\hfill  &
\hfill 2.27\hfill  & \hfill -0.0482\hfill  \cr
\+ 25\% Highest IBP = ``Ellipticals"\hfill &\hfill\hskip 1.cm 1.126\hfill  &
\hfill 2.25\hfill  & \hfill -0.0455\hfill  \cr
\+ 50\% Lowest IBP = ``Spirals"\hfill &\hfill\hskip 1.cm 1.193\hfill & \hfill
2.26\hfill  &	\hfill -0.0671\hfill  \cr
}$$
\vfill\eject

\hsize=5.25truein
\hoffset=2.45truecm
\centerline{FIGURE CAPTION}
\medskip

\item{Fig. 1--}
A slice through the hydrodynamic simulations with
co-moving dimensions
$80 h^{-1}\mpc \times 80 h^{-1} \mpc \times 15 h^{-1} \mpc$.
(a):  sample of 1/2 of the
spirals (youngest 50\%)
at z=0,
(b):  S0's (intermediate 25\%) at  z=0,
(c):  ellipticals (oldest 25\%) at z=0,
(d):  ellipticals at z=3.6.  At the present epoch,
spirals are seen on a network
of walls and filaments, while
ellipticals congregate more in clusters.
The picture at z=3.6 shows that these ellipticals
originally formed on a network of
walls and filaments just like the spirals and then
later drained gravitationally
onto the clusters.  The S0's form an intermediate population.

\item{Fig. 2--}
Genus curve for the dark matter
(CDM $\Omega=1$, $h=0.5$) in the hydrodynamical
simulations at the present epoch.
In this diagram and in those that follow, the
total genus for the simulation cube
$(80 h^{-1}\mpc)^3$ is shown, which is equal to
the number of ``donut" holes minus the number of isolated regions in the
smoothed density-contour surfaces.  The volume fraction
$f_V$ on the low-
density side of the contour surface is indicated by the parameter
$\nu$:
$f_V = (2.5\%, 16\%, 50\%, 84\%, 97.5\%)$
for $\nu = (-2, -1, 0, 1, 2)$ respectively.
The data is shown as a solid curve.
The 1 $\sigma$ error bars shown are calculated
using the standard deviation of the mean genus for the entire simulation cube
derived from the genus values from the 8 independent subcubes of
$(40 h^{-1} \mpc)^3$
which together make up the entire simulation cube of
$(80 h^{-1} \mpc)^3$.
The best fit random-phase curve
$g(\nu) \propto (1 - \nu^2) \exp (- \nu^2/2)$
is shown as a
dashed line.  The overall genus curve for the dark matter is approximately
random phase, reflecting the topology of the initial conditions.  The median
density contour is spongelike
[$g (\nu = 0) > 0]$.  The smoothing length is
$\lambda = 8h^{-1}\mpc=800\kms$.

\item{Fig. 3--}
The genus curve for all galaxies in the hydrodynamic simulations at the
present epoch, plotted as in fig. 2.  Again the genus curve is consistent with
the predicted random-phase curve inherited from the initial conditions.  There
is a slight shift to the left
(toward a meatball topology) characteristic of that
seen in many biased CDM models (cf. Gott et al., 1989).

\item{Fig. 4--}
Figure 4a shows
the genus curve for ellipticals (the first 25\% of galaxies formed) from the
hydrodynamic simulations at the present epoch.
Figure 4b shows
the genus curve for S0 galaxies (the second 25\% of galaxies formed)
from the hydrodynamic simulations at the present epoch.  This population is
intermediate between ellipticals and spirals.
Similarly, Figure 4c shows the genus curve for spirals (the 50\% of galaxies
most recently formed).
The ellipticals (Fig. 4a) show a genus curve that is shifted to the left
(toward the meatball topology) relative to that of the spirals (Fig. 4c).

\item{Fig. 5--}
Figure 5a shows the
$\nu=-1$ density-contour surface is shown for ellipticals in the simulation
cube.  This surface encloses the 16\% of the volume which is lowest density.
For comparison,
Figure 5b shows
the same $\nu=-1$ density contour surface is shown for
the spirals.
Because of the meatball shift to the left for the ellipticals, this
particular contour has a more spongelike topology than for spirals.
Fig. 4a,c
shows that
$g(-1)\approx 3$
for ellipticals while
$g(-1) \approx 2$
for spirals.  In the figure
this difference of $+1$
in the elliptical contour surface is caused by the tube
shown in 5a connecting the two voids at the bottom and on the right, which is
absent in the spiral contour surface in 5b.  One more connection (one more
``donut" hole) means a difference in genus of $+1$.
Figures (5c,d) show
the $\nu = 0$ median density contour surfaces for ellipticals and spirals
are shown in figures 5c and 5d respectively; both are spongelike.

\item{Fig. 6--}
Density-contour surfaces can be labeled by either the fraction of the volume
contained on their low-density side
$f_V$ or by the mass fraction in the smoothed
mass distribution
$f_M$ contained on their low-density side.  This graph shows
the relation between
$f_V$ and $f_M$ for dark matter, all galaxies, ellipticals, and spirals.

\item{Fig. 7--}
Figure 7a shows
the genus curve for elliptical galaxies plotted as a function of $f_M$.
Figure 7b shows that for spirals.
 The genus curve for ellipticals becomes negative
(signifying isolated regions) for
$f_M > 0.58$ while for spirals the genus
becomes negative for
$f_M > 0.70$.  This means that 42\% of the smoothed mass
distribution of ellipticals is in isolated clusters, while only 30\% of the
smoothed mass distribution of spirals is in isolated clusters.

\item{Fig. 8--}
Results at the present epoch from a standard biased N-body simulation for
comparison.
Figure 8a shows genus curve for the 25\% highest biased particles drawn from
the initial
conditions,
these will be the galaxies that form first - the ellipticals.
Figure 8b shows genus curve for the 25\% next highest biased particles drawn
from the
initial conditions, these will form next - the S0's.
Figure 8c shows genus curve for the 50\% lowest biased particles
drawn from the initial
conditions, these will form last - the spirals.
There is no systematic shift of
the elliptical curve to the left relative to the spirals as was seen in the
hydrodynamical simulations.

\item{Fig. 9--}
Genus curves for ellipticals (9a) and spirals (9b),
respectively, plotted as a
function of $f_M$ for the biased
N-body simulation (to compare with figs. 7a and
b for the hydrodynamical simulations).  In 9a, the genus becomes negative at
$f_M = 0.68$, while in 9b the genus becomes negative at
$f_M = 0.71$.  Thus, the
fraction of the smoothed mass distribution in ellipticals which is in isolated
clusters (32\%) differs from the fraction of the smoothed mass distribution in
spirals which is in isolated clusters (29\%) by only a small amount (3\%).

\item{Fig. 10--}
Figure 10a shows
the genus curve for elliptical galaxies plotted as a function of $f_M$
with a smoothing length of $560\kms$.
Figure 10b shows that for spirals.
As expected,
the elliptical and spiral genus curves are even
more differentiated from each other.
At this smoothing length $52\%$ of the smoothed mass distribution
in the ellipticals is in isolated clusters whereas only
$37\%$ of the smoothed mass distribution
in the spirals is in isolated clusters (a difference of $15\%$).

\vfill\eject\end